\documentclass[Afour,sagev,times]{sagej}
\usepackage{lipsum}
\usepackage{moreverb,url}
\usepackage{epsfig} 
\usepackage{multicol, blindtext}
\usepackage[caption=false]{subfig}
\usepackage{amsmath}
\usepackage{float}
\usepackage{commath}
\usepackage[superscript]{cite}
\usepackage{placeins}
\makeatletter
\renewcommand\@biblabel[1]{#1.}
\makeatother
\usepackage[colorlinks,bookmarksopen,bookmarksnumbered,citecolor=red,urlcolor=red]{hyperref}

\newcommand\BibTeX{{\rmfamily B\kern-.05em \textsc{i\kern-.025em b}\kern-.08em
T\kern-.1667em\lower.7ex\hbox{E}\kern-.125emX}}

\begin{document}

\runninghead{Panda and Warrior}

\title{Modelling the Pressure Strain Correlation in turbulent flows using Deep Neural Networks}
\author{J. P. PANDA AND H. V. WARRIOR} 

\affiliation{Department of Ocean Engineering and Naval Architecturte\\Indian Institute of Technology, Kharagpur, India}

\corrauth{J P Panda, IIT Kharagpur, INDIA}

\email{jppanda@iitkgp.ac.in}

\begin{abstract}
The pressure strain correlation plays a critical role in the Reynolds stress transport modelling. Accurate modelling of the pressure strain correlation leads to proper prediction of turbulence stresses and subsequently the other terms of engineering interest. However, classical pressure strain correlation models are often unreliable for complex turbulent flows. Machine learning based models have shown promise in turbulence modeling but their application has been largely restricted to eddy viscosity based models. In this article, we outline a rationale for the preferential application of machine learning and turbulence data to develop models at the level of Reynolds stress modeling. As the illustration We develop data driven models for the pressure strain correlation for turbulent channel flow using neural networks. The input features of the neural networks are chosen using physics based rationale. The networks are trained with the high resolution DNS data of turbulent channel flow at different friction Reynolds numbers. The testing of the models are performed for unknown flow statistics at other friction Reynolds numbers and also for turbulent plane Couette flows. Based on the results presented in this article, the proposed machine learning framework  exhibits considerable promise and may be utilized for the development of accurate Reynolds stress models for flow prediction.
\end{abstract}


\maketitle

\section{Introduction}
Reliable Computational Fluid Dynamics(CFD) \cite{versteeg2007introduction} entails the modelling and prediction of fluid flows of engineering interest starting from flows in simple channels \cite{moser1999direct} to environmental flows \cite{mitra2019effects}, where the length scale is very high \cite{mitra2020experimental}. The building block of such CFD codes are often the turbulence models \cite{pope2001turbulent}. Accurate modelling of turbulence, will lead to improved prediction of fluid flows involving complex strain or rotational fields and complex flows involving adverse pressure gradient and stream line curvature\cite{celic2006comparison}. 

The basic CFD treatment of turbulence can be mainly classified as eddy viscosity models, Reynolds stress transport models\cite{panda2018representation,panda2020review}, Large eddy simulations and direct numerical simulations\cite{pope2001turbulent}. The eddy viscosity models have lowest level of accuracy. The major drawback of such models is the Boussinesq eddy viscosity hypothesis, in which the eddy viscosity is defined in terms of the local strain fields, which may focus on empiricism. Although these models are less accurate, but the computational expanse associated with such models is lower, since these employ only two equations for the prediction of turbulence stresses(one for turbulence kinetic energy and another for dissipation). Because of their simplicity those are largely used in industrial applications involving larger flow domain and complex operating conditions. In contrast to eddy viscosity models, the cost of LES and DNS simulations are very high, so those are not suitable for industrial flows. 

The current emphasis of turbulence modelling community have been shifted towards Reynolds stress models\cite{mishra2017toward,mishra2010pressure} with increase in computational facility. The cost and accuracy of Reynolds stress transport models lies between eddy viscosity and LES approaches. The Reynolds stress models have transport equations for each component of Reynolds stress, from which the turbulent stress field can be directly obtained. The detailed structure of the Reynolds stress transport equation will be discussed in the next section. The Reynolds stress transport models mainly employs models for the pressure strain correlation. The pressure strain correlation model incorporates complex flow physics resulting from the turbulence/turbulence or turbulence/mean-strain interactions into the modelling basis. Several researchers has provided different formulations separately for the slow\cite{sarkar1990simple,chung1995nonlinear,panda2017improved} and rapid pressure strain correlations\cite{panda2018representation}. Most standard complete pressure strain correlation models are \cite{mishra2017toward,speziale1991modelling,launder1975progress}. These are applied to many turbulent flow simulations but the major drawback of such models are the model coefficients, which are calibrated by using very limited number of cases of experimental or DNS results, which raises questions on the universality of the model. These models are often unreliable for complex problems involving flows of real life engineering application. 

In the recent decade there has been a marked increase in the development and adoption of Machine Learning (ML) algorithms. This has been bolstered by the availability of open-source tools for their application, the increment in computational resources and the availability of data. Such machine learning algorithms have found applications in physical sciences, such a high energy physics, material sciences, etc. To this end there has been a shift in focus from the use of empirical approaches toward formulating turbulence models, to a purely data driven approach to produce these models. Several CFD researchers\cite{renganathan2020machine,maulik2018data} and turbulence modellers have adopted machine learning methodologies for improved flow predictions either by developing surrogate fluid dynamics models or directly modelling the terms in the transport equations\cite{zhang2018machine,zhang2019application,lav2019framework,PhysRevFluids.5.114604,pandey2020perspective, tracey2015machine}.  In surrogate fluid dynamic models the modellers use results obtained from computational fluid dynamics simulations to train the machine learning models and later use the ML model to predict flow parameters for other operating conditions, e.g. \cite{maulik2020turbulent} used results obtained from SA model simulations for a backward facing step to train ML models and later they used the trained ML model to predict flow properties at different flow velocities and step heights. Although surrogate models are suitable for fast prediction of flow parameters, but those should not completely replace the physics based CFD models. Sekar et al.\cite{sekar2019fast} used both convolutional neural network and artificial neural network to predict the hydrodynamics parameters such as drag, lift and pressure coefficients of an airfoil. They employed CNN to obtain the geometrical parameters of the aerofoil. In conjunction with the airfoil geometrical parameters they used Reynolds number and angle of attack data to train the model using ANN. Hui et al.\cite{hui2020fast} used deep learning methodologies for fast prediction of pressure distribution over airfoils. Deng et al. \cite{deng2019time} used long short-term memory based artificial intelligence framework for time-resolved turbulent flow reconstruction from discrete point measurements and non-time-resolved
particle image velocimetry measurements.  

Researchers are also using domain knowledge to enforce physics based constrains on such machine learning models. These vary from symmetry requirements, to conservation of mass, translation and Galilean invariance, etc. These are termed as physics informed machine learning models of turbulence\cite{wang2017physics}. Wang et al. \cite{wang2017physics} presented a comprehensive framework for augmenting turbulence models with physics-informed machine learning, with a complete work flow from identification of input features to  prediction of mean velocities. They predicted the linear and non-linear parts of the Reynolds stress tensor separately. Zhu et al.  \cite{zhu2019machine} constructed a mapping function between turbulence mean viscosity and mean flow variables by using neural networks(A detailed discussion on neural networks will be provided in subsequent sections) and they had completely replaced the original partial differential equation models. They used data generated by Spalart- Allmaras(SA) model as training data and used a radial basis function neural network for the development of the machine learning model. Fang et al. \cite{fang2020neural} used artificial neural networks(ANN) to develop machine learning models for the anisotropic Reynolds stress tensors and also proposed several modifications to the simple multilayer perception for incorporating no-slip boundary condition, Reynolds number and non-local effects. They used high fidelity turbulent channel flow DNS data \cite{lee2015direct} at different Reynolds numbers for training their model. Yin et al. \cite{yin2020feature} proposed a selection criteria for the input features of a neural network based on physical and tensor analysis. The first step of such approach is to search for tensors and vectors upon which the Reynolds stress has dependency and to construct a tensor basis for representing the Reynolds stress. Ling et al.\cite{ling2016reynolds} proposed a modified deep network(Tensor basis neural network) and learned a model for the Reynolds stress anisotropy form the high fidelity simulation data. The modified neural network has a multiplicative layer with an invariant tensor basis to embed Galilean invariance \cite{pope2001turbulent} into the predicted anisotropy tensor. Singh et al.\cite{singh2017machine} employed neural networks and field inversion techniques for introducing correction factors in the SA model. They employed the improved SA model to predict the separated flow over airfoils. Parish and Duraisamy\cite{parish2016paradigm} also used similar approach to modify the turbulence kinetic energy equation. weatheritt et al. \cite{weatheritt2016novel} used gene expression programming for algebraic modification of the RANS stress strain relationship. Mathematical model for the tensors are created using high fidelity data and uncertainty measures. Weatheritt et al.\cite{weatheritt2017development} used similar approach to develop algebraic stress models,  the model was created hybrid RANS/LES flow field data. Taghizadeh at al.\cite{taghizadeh2020turbulence} provided series of guidelines for the alteration of coefficients of turbulence models in machine learning assisted turbulence modelling, such that the characteristics of the real physics based models are preserved.

The general methodology utilized in most of these studies pertains to using large corpora of high fidelity data from DNS or LES simulations along with a machine learning algorithm, such as deep neural networks or random forests. The machine learning model is trained on the learning dataset to infer optimal coefficients for the closure of the turbulence model. The form of the turbulence model pertains to classical 2-equation based eddy-viscosity based models (EVM) or Algebraic Reynolds Stress Models (ARSM). While such approaches have shown success, this methodology may be impaired by the dissonance between the fidelity of the data and the maximum potential fidelity of the baseline model form utilized. For instance, the data from DNS studies reflects high degrees of anisotropy in the turbulent flow field. However any eddy viscosity based model, even with optimal coefficients inferred using machine learning, is incapable of capturing high degrees of turbulence anisotropy due to the nature of the linear eddy viscosity hypothesis inherent to the model \cite{mishra2019estimating}. This eddy viscosity hypothesis states that the turbulence anisotropy is a function of the instantaneous mean strain rate and thus must lie on the ``plane strain" manifold of the barycentric triangle \cite{edeling2018data}. Consequently the anisotropy information from the high fidelity data is rendered ineffectual due to the form of the baseline model. Similarly the high fidelity data subsumes the complex dependency of the turbulent statistics on the mean rate of rotation and the effects of streamline curvature. However in the eddy viscosity based models the Reynolds stresses are only dependent on the mean rate of strain \cite{pope2001turbulent}. Thus these rotational effects are unusable due to the form of the baseline model \cite{mishra2019linear}. Similarly Algebraic Reynolds Stress Models presuppose that the convective and diffusive fluxes in the turbulent flow are negligible, or the flow is source dominated \cite{gatski1993explicit}. This is an extremely restrictive assumption and may not be valid for most turbulent flow datasets.  In this context it may be advisable to use a different baseline model formulation that is expressive and flexible enough to take advantage of the information in the high fidelity data. The Reynolds Stress Modeling approach offers such an alternative. Instead of presupposing any form of a linear relationship between the mean gradients and the Reynolds stress tensor, the Reynolds stress modeling approach utilizes the Reynolds Stress Transport Equations to generate evolution equations for each component of the turbulent anisotropy tensor. This direct and explicit computation of the evolution of each term of the turbulent anisotropy leads to better representation of the state of anisotropy in turbulence. Reynolds stress modeling approach has the ability to account for the directional effects of the Reynolds stresses and the complex interactions in turbulent flows. These models can represent complex turbulent physics in limiting states such as the return to isotropy of turbulence found in decaying turbulent flows and the dynamics of turbulence at the Rapid Distortion Limit where the turbulent flow behaves similar to an elastic medium. Due to the explicit modeling of different turbulent transport processes, Reynolds stress models can account for the complex effects of flow stratification, buoyancy effects, streamline curvature, etc. Consequently in a machine learning framework, using Reynolds Stress Models as the baseline models may enable the utilization of a significantly higher degree of physics and information inherent in the high fidelity data. However at present there has been little research to develop the potential of the Reynolds Stress Modeling approach using machine learning approaches. This is the central novelty of this investigation. 

The Reynolds Stress Modeling approach depends on the development of surrogate models to represent different turbulence transport processes. These include turbulent transport, rotational effects, rate of dissipation and the pressure strain correlation. While reliable models for all these terms are essential, the modeling of the pressure strain correlation term has been a long standing challenge in turbulence modeling. The pressure strain correlation term represents physics responsible for the transfer of energy between different components of the Reynolds stress tensor \cite{mishra2014realizability}. It is responsible for the non-local interactions in turbulent flows, the initiation of instabilities in rotation dominated flows, the return to isotropy observed in decaying flows, etc \cite{mishra2013intercomponent}. While classical models have been developed for the pressure strain correlation term, such physics driven models have many limitations in their ability to account for streamline curvature effects, realizability requirements, their performance in complex engineering flows \cite{mishra2017toward}. In this context this investigation focuses on the utilization of machine learning approaches for the formulation of data driven models for the pressure strain correlation. In this article, we have modelled the pressure strain correlation for turbulent channel flow using deep neural networks. The input features of the neural network were chosen using physics based approaches. The high fidelity DNS data of turbulent channel flow at different friction Reynolds numbers are used to train the neural network. A set of unknown cases of turbulent channel flow at different friction Reynolds numbers are used to test the predictive capability of the neural network model. An important requirement from machine learning models is generalizability. Here, the model is expected to perform well not only in the cases that were included in its training, but must also perform well in similar cases that were not included in its training. To evaluate the generalizability characteristics of the deep learning models, an additional dataset of turbulent Couette flow at a particular friction Reynolds number is also used to check their predictive capability. 
\section{Reynolds stress transport modelling framework}
The building block of Reynolds stress transport models are the Reynolds stress transport equations. This set of equations outline the evolution of different components of the Reynolds stress tensor in a turbulent flow. This evolution is affected by different transport processes in turbulence, that are represented by different terms in the set of equations.
The Reynolds stress transport equation has the form:
\begin{equation}
\begin{split}
&\partial_{t} \overline{u_iu_j}+U_k \frac{\partial \overline{u_iu_j}}{\partial x_k}=P_{ij}-\frac{\partial T_{ijk}}{\partial x_k}-\eta_{ij}+\phi_{ij},\\
&\mbox{where},\\ 
& P_{ij}=-\overline{u_ku_j}\frac{\partial U_i}{\partial x_k}-\overline{u_iu_k}\frac{\partial U_j}{\partial x_k},\\
&\ T_{ijk}=\overline{u_iu_ju_k}-\nu \frac{\partial \overline{u_iu_j}}{\partial{x_k}}+\delta_{jk}\overline{ u_i \frac{p}{\rho}}+\delta_{ik}\overline{ u_j \frac{p}{\rho}},\\
&\epsilon_{ij}=-2\nu\overline{\frac{\partial u_i}{\partial x_k}\frac{\partial u_j}{\partial x_k}}  \\
&\phi_{ij}= \overline{\frac{p}{\rho}(\frac{\partial u_i}{\partial x_j}+\frac{\partial u_j}{\partial x_i})}\\
\end{split}
\end{equation}
$P_{ij}$ denotes the production of turbulence, $T_{ijk}$ is the diffusive transport, $\epsilon_{ij}$ is the dissipation rate tensor and $\phi_{ij}$ is the pressure strain correlation. The pressure fluctuations are governed by a Poisson equation:
\begin{equation}
\frac{1}{\rho}{\nabla}^2(p)=-2\frac{\partial{U}_j}{\partial{x}_i}\frac{\partial{u}_i}{\partial{x}_j}-\frac{\partial^2 u_iu_j}{\partial x_i \partial x_j}
\end{equation}
The fluctuating pressure term is split into a slow and rapid pressure term $p=p^S+p^R$. Slow and rapid pressure fluctuations satisfy the following equations
\begin{equation}
\frac{1}{\rho}{\nabla}^2(p^S)=-\frac{\partial^2}{\partial x_i \partial x_j}{(u_iu_j-\overline {u_iu_j})}
\end{equation}
\begin{equation}
\frac{1}{\rho}{\nabla}^2(p^R)=-2\frac{\partial{U}_j}{\partial{x}_i}\frac{\partial{u}_i}{\partial{x}_j}
\end{equation}
The slow pressure term accounts for the non-linear interactions (turbulence-turbulence interactions) in the fluctuating velocity field and the rapid pressure term accounts for the linear interactions(mean strain-turbulence interactions). The pressure strain correlation can be modeled using rational mechanics approach. The rapid term is modelled as
\cite{pope2001turbulent}

\begin{equation}
\phi_{ij}^R=4k\frac{\partial{U}_l}{\partial{x_k}}(M_{kjil}+M_{ikjl})
\end{equation}
where, 
\begin{equation}
M_{ijpq}=\frac{-1}{8\pi k}\int \frac{1}{r} \frac {\partial^2 R_{ij}(r)}{\partial r_p \partial r_p}dr
\end{equation}
where, $R_{ij}(r)=\langle u_i(x)u_j(x+r) \rangle$
For homogeneous turbulence the complete pressure strain correlation can be written as
\begin{equation}
\phi_{ij}=\epsilon A_{ij}(b)+kM_{ijkl}(b)\frac{\partial\overline {v}_k}{\partial{x_l}}
\end{equation}
The most general form of slow pressure strain correlation is given by
\begin{equation}
\phi^{S}_{ij}=\beta_1 b_{ij} + \beta_2 (b_{ik}b_{kj}- \frac{1}{3}II_b \delta_{ij})
\end{equation}
Established slow pressure strain correlation models including the models of \cite{sarkar1990simple} use this general expression. Considering the rapid pressure strain correlation, the linear form of the model expression is
\begin{equation}
\begin{split}
&\frac{\phi^{R}_{ij}}{k}=C_2 S_{ij} +C_3 (b_{ik}S_{jk}+b_{jk}S_{ik}-\frac{2}{3}b_{mn}S_{mn}\delta_{ij})+  
\\&C_4 (b_{ik}W_{jk} + b_{jk}W_{ik})
\end{split}
\end{equation}
Here $b_{ij}=\frac{\overline{u_iu_j}}{2k}-\frac{\delta_{ij}}{3}$ is the Reynolds stress anisotropy tensor, $S_{ij}$ is the mean rate of strain and $W_{ij}$ is the mean rate of rotation. Rapid pressure strain correlation models use this general expression. 
The most widely used Reynolds stress transport model is proposed by \cite{speziale1991modelling}, which has the form:

\begin{equation}
\begin{split}
& \phi_{ij}^{(R)}= C_1 \epsilon b_{ij} + C_2 \epsilon (b_{ik}b_{kj}-\frac{1}{3} b_{mn}b_{mn}\delta_{ij}) + \\ &
C_3 K S_{ij}+ C_4K(b_{ik} S_{jk}+b_{jk} S_{ik}-2/3b_{mn} S_{mn}\delta_{ij})\\ & +C_5 K (b_{ik} W_{jk}+b_{jk} W_{ik})
\end{split}
\end{equation}
\begin{figure}
\centering
 \subfloat[]{\includegraphics[width=0.45\textwidth]{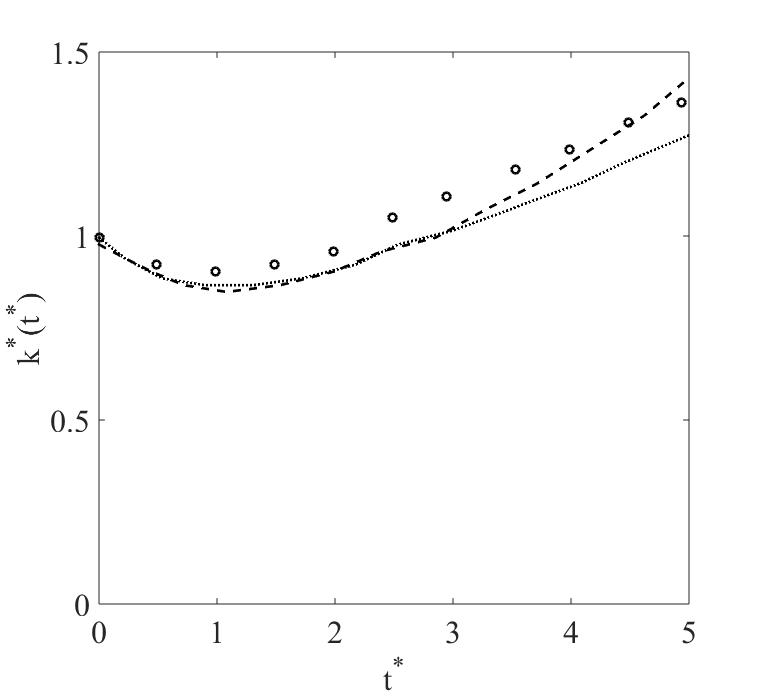}}\\
 \subfloat[]{\includegraphics[width=0.45\textwidth]{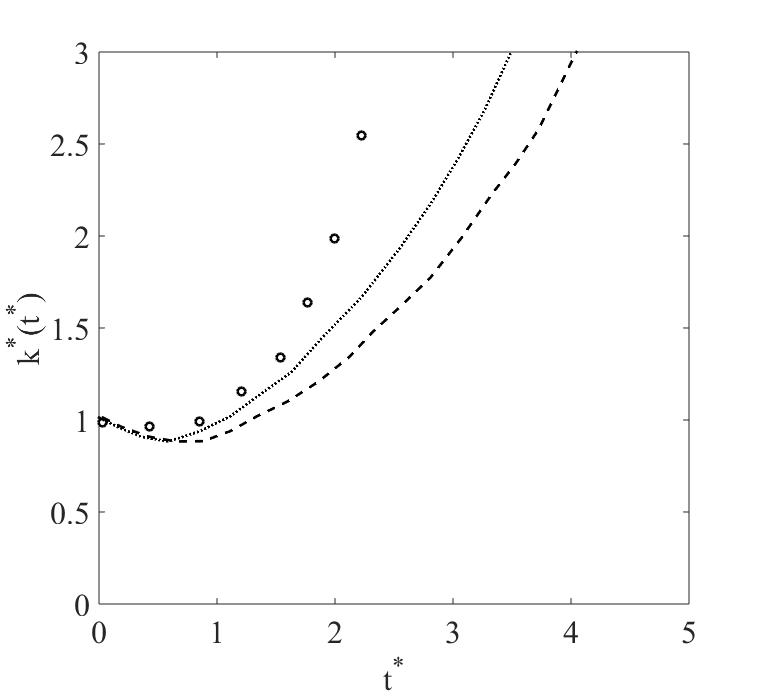}}\\
 \subfloat[]{\includegraphics[width=0.45\textwidth]{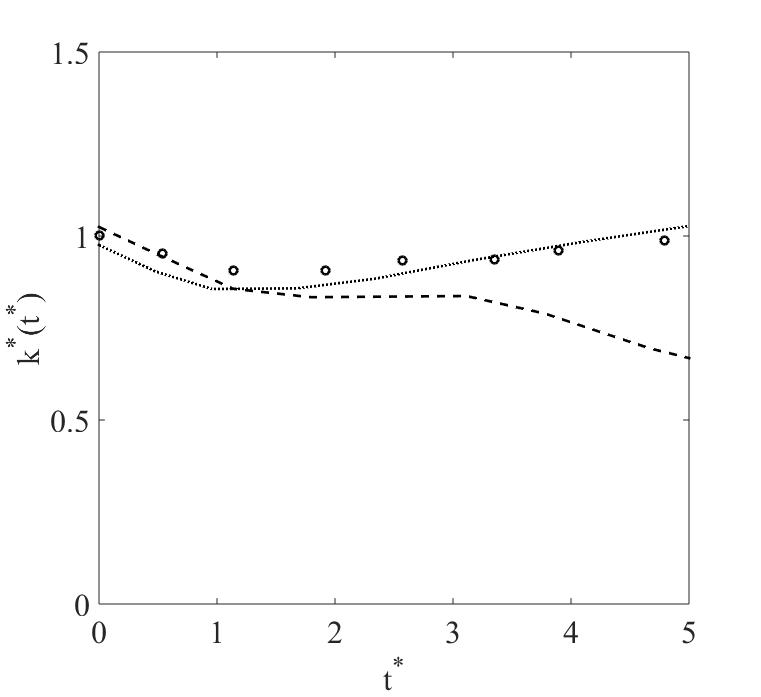}}
 \caption{Evolution of turbulence kinetic energy for rotating shear flows, a) $W/S=0$, b) $W/S=0.25$ and c) $W/S=0.5$, Symbols LES data, dashed lines predictions of LRR model, dotted lines predictions of SSG model.\label{fig:11}}
\end{figure}
The expressions for these rapid and slow pressure strain correlation models have physical significance where different terms represent interactions and effects from different physical mechanisms. For instance, in the slow pressure strain correlation expression, both the terms attempt to produce the return to isotropy behavior of decaying turbulence. The first term, $\beta_1 b_{ij}$, represents a linear decay and leads to evolution along linear paths on the anisotropy invariant map. The second term, $b_{ik}b_{kj}- \frac{1}{3}II_b \delta_{ij}$ represents nonlinear effects of the turbulent anisotropy on the decay. This produces nonlinear paths of decay on the anisotropy invariant map. Similar reasoning can be made about the terms in the rapid pressure strain correlation model. While carrying out the data driven modeling, the input features for the neural network model will be selected based on the above tensor representation and their implied physical processes.
\begin{figure*}[ht]
\centering
\captionsetup{justification=centering}
\includegraphics[width=0.6\textwidth]{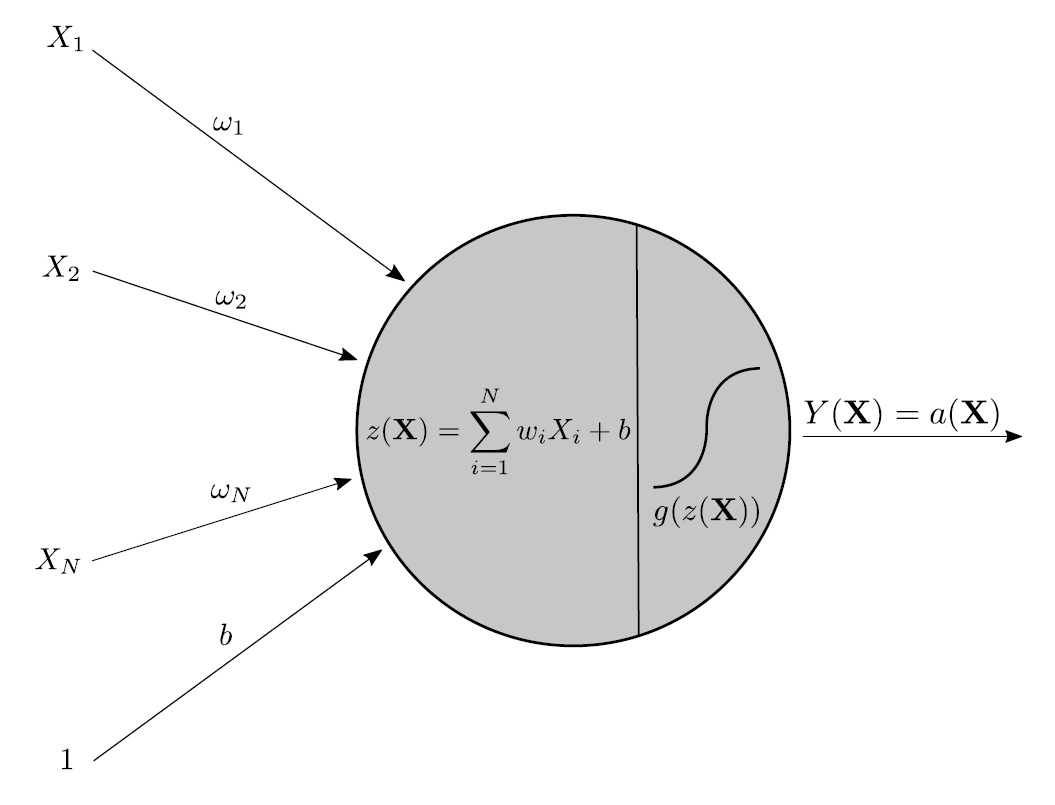}
\caption {The structure of an artificial neuron in a neural network.\label{fig:1}}
\end{figure*}


\section{Limitations of the pressure strain correlation models}
The pressure strain correlation model coefficients are calibrated with very few  cases of turbulent flow data sets. So, when the models are tested for other unknown flow prediction problems, those produces unrealistic results. Another, limitation in Reynolds stress transport modelling approach with models for pressure strain correlations is that, they can not accurately replicate the non-local nature of flow, since the models of pressure strain correlation has only local terms in the modelling basis. As can be seen this is not a limitation of the data but of the modeling form selected. A more flexible and expressive model form may be able to infer additional physics based information from the data and lead to better predictions. Based on the universal approximation theorem, neural networks can learn arbitrarily complex functions from data contingent upon the selection of proper hyperparameters \cite{hornik1991approximation}. Thus the utilization of such a machine learning algorithm may be justified in these circumstances. 

From analysis of various established pressure strain correlation models it is observed that, there is a mismatch between the trend of evolution of turbulence kinetic energy with LES data in rotation dominated flow fields. A critical comparison of established pressure strain correlation models for turbulence kinetic energy evolution is presented in fig.\ref{fig:11}. The symbols in the fig.\ref{fig:11} represent the LES results of\cite{bardino1983improved}. Fig. \ref{fig:11}a, b and c represent the evolution of turbulence kinetic energy for three different $W/S$ ratio. Although at $W/S=0$ the predictions of turbulent kinetic energy is matching with the LES results but there is a significant mismatch between the model predictions and the LES results for higher magnitude of $W/S$ ratio. We observe that there are critical limitations in the pressure strain correlation models under use presently.        
  
\section{Turbulent channel flow}
This case of flow consists of fluid in between two infinite parallel plates in x-z plane. The plates are situated at y=0 and y=2h. The flow is basic pressure driven with known pressure gradients. The three components of velocity are u, v and w respectively. The velocity components are functions of x, y, z and t respectively, where x, y, z are space coordinates and t is time. The friction Reynolds number for the channel flow can be defined as $Re_{\tau}=\frac{u_{\tau}h}{\nu}$. Where, $u_{\tau}=\frac{\tau_{wall}}{\rho}$ is the friction velocity. The fluid density and kinematic viscosity are $\tau$ and $\nu$ respectively. A non-dimensional wall from the wall can be written as $y^+=\frac{u_{\tau}y}{\nu}$.   

Turbulent velocity can be decomposed in mean and fluctuating velocities. The averaging of the velocity field in the Navier-stokes equation resulted in the Reynolds averaged Navier-Stokes equations(RANS). In the RANS equations there are 4 equations and 10 unknowns. Taking moment of RANS equations the Reynolds stress transport equations can be derived. In the Reynolds stress transport equations the pressure strain correlation term is the most important term, that needs to be modelled, for incorporating complex flow physics in to the modelling basis. The pressure strain correlation has direct relation with turbulence dissipation($\epsilon$), Reynolds stress anisotropy($b_{ij}$), turbulence kinetic energy($k$) and velocity gradient($S_{ij}$). These terms can be defined as follows:

\begin{equation}
\begin{split}
& \epsilon=\nu \overline{\frac{\partial{u}_i}{\partial{x_k}} \frac{\partial{u}_i}{\partial{x_k}}}
\end{split}
\end{equation}  
\begin{equation}
\begin{split}
& b_{ij}=\overline{\frac{u_i u_j}{2k}}- \frac{\delta_{ij}}{3}
\end{split}
\end{equation} 

\begin{equation}
\begin{split}
&S_{ij}=\frac{1}{2}(\frac{\partial{U}_i}{\partial{x_j}}+\frac{\partial{U}_j}{\partial{x_i}})
\end{split}
\end{equation} 

\section{Neural Networks for modelling of the pressure strain correlation}
\begin{figure*}[ht]
\centering
\captionsetup{justification=centering}
\includegraphics[width=0.6\textwidth]{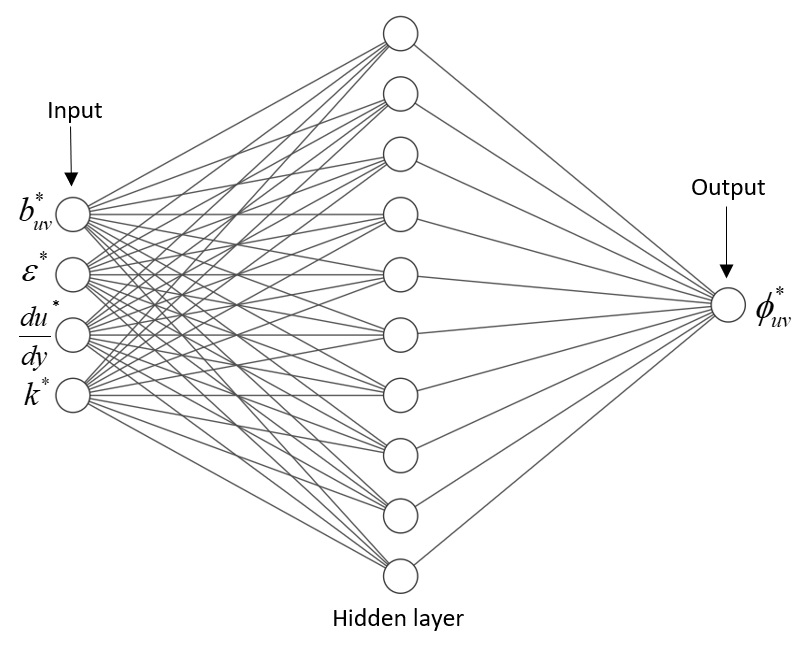}
\caption {Diagram of the neural network(NN1) for the turbulent channel flow. The inputs are the Reynolds stress anisotropy, dissipation, velocity gradient, and turbulence kinetic energy and the output is the pressure strain term.The FCFF has 1 layer with 10  neurons.\label{fig:2}}
\end{figure*}

\begin{figure*}[ht]
\centering
\captionsetup{justification=centering}
\includegraphics[width=0.7\textwidth]{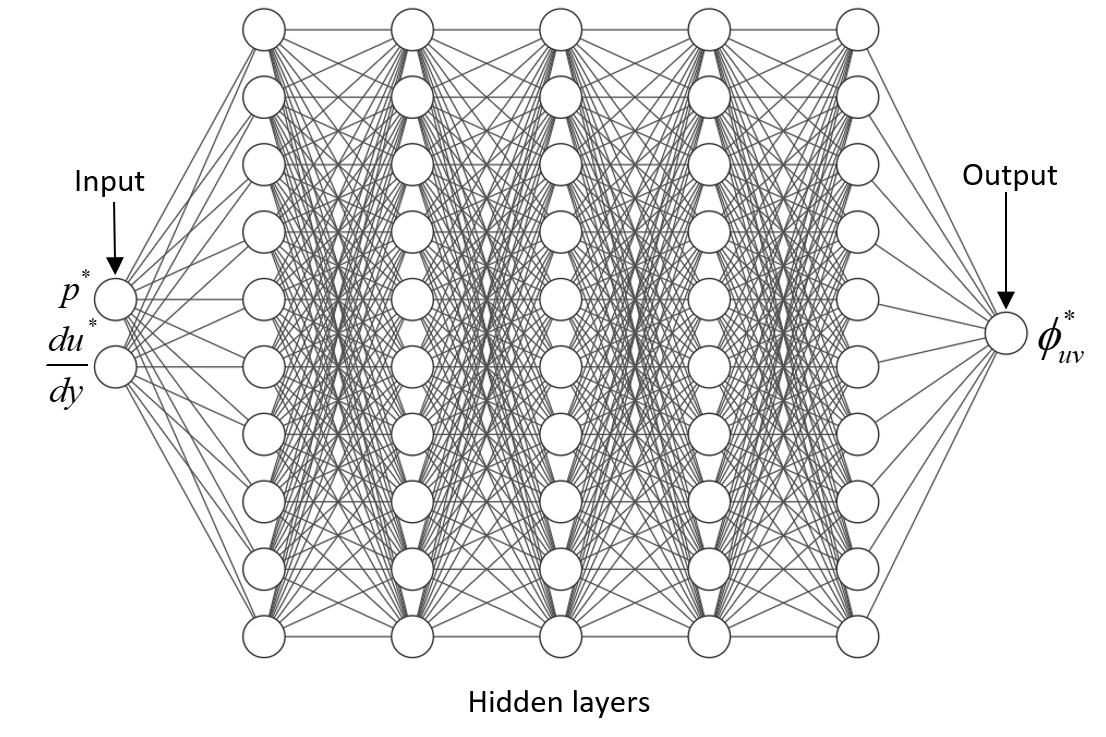}
\caption {Diagram of the neural network(NN2) for the turbulent channel flow. The inputs are the pressure and velocity gradient and the output is the pressure strain term. The FCFF has 5 layers with 10 neurons in each layer.\label{fig:3}}
\end{figure*}

\begin{figure*}[ht]
\centering
\captionsetup{justification=centering}
\subfloat[$NN1$]{\includegraphics[width=0.5\textwidth]{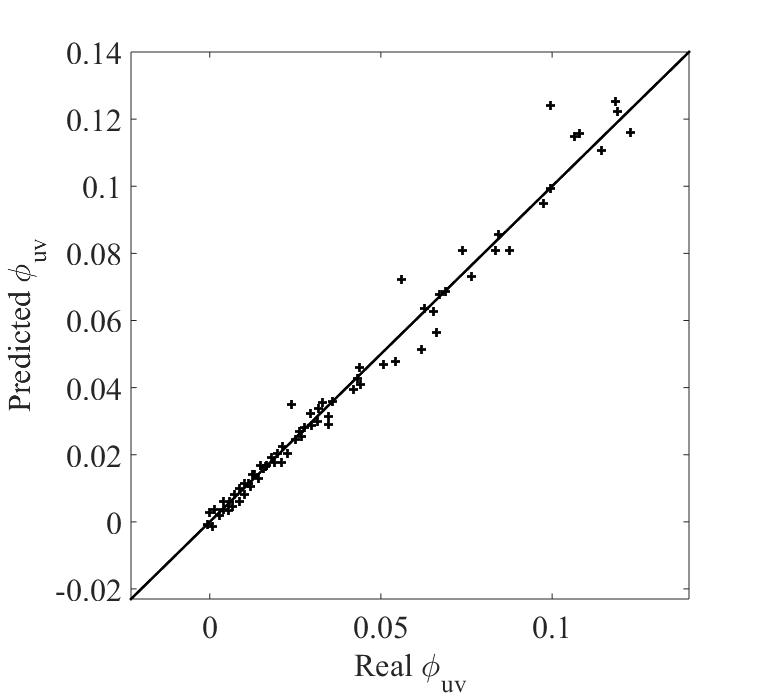}}
 \subfloat[$NN2$]{\includegraphics[width=0.5\textwidth]{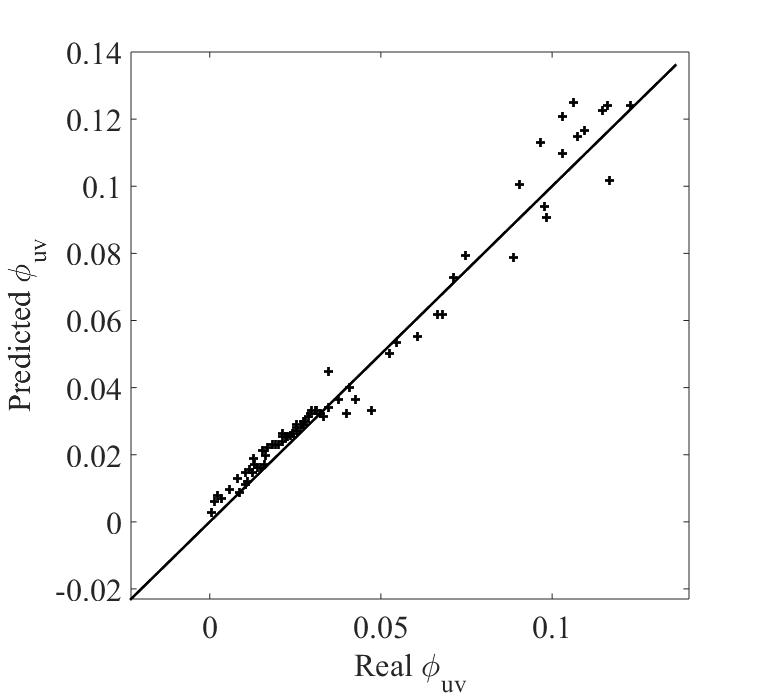}}
 \caption{Actual values vs predicted values of the pressure strain correlation for $Re_\lambda=5200$.\label{fig:4}}
\end{figure*}
Neural networks are a supervised machine learning algorithm that can be utilized for regression and classification tasks. There are various types of neural networks with different architectures and neuron connection forms(the structure of an artificial neuron is shown in fig. \ref{fig:1}), e.g. fully connected neural networks(FCNN), convolutional neural networks(CNN),  recurrent neural networks(RNN), etc. In this work, feed forward FCNN is considered, where the inputs are correlated with the outputs, with the transformation of inputs through non-linear activation functions. The layers between input($z_0\in R^{n_0}$) and output layers($z_L\in R^{n_L}$) are called hidden layers. A neural network with more than one hidden layer is called deep neural network(DNN). Two adjacent layers in a DNN are mathematically connected as,
\begin{equation}
\begin{split}
&z_l=\sigma_l(W_l^T z_{l-1}+b_l), 
\end{split}
\label{eq:14}
\end{equation}
in the above equation, $W_l\in R^{{n_{l-1}}\times n_{l}}$ and $b_l\in R^{n_l}$ are the weight matrix and bias vector, the index of the layers is represented as subscript $l$. $\sigma_l(.)$ is the activation function. In this work non-linear activation function is used. The function of non-linear activation function is to add non-linear real world properties to the artificial neural network. If the activation function is not applied, a simple linear function will be resulted as output, that is undesirable. The neural network without any activation function will act as linear regression. The neural network with non-linear activation functions behave as an universal approximator. There are different types of activation functions are available for use in neural networks. Most widely used activation functions are logistic, hyperbolic Tangent and  ReLU(Rectified Linear Units). More information on such activation functions is available in \cite{witten2002data}.

The output of the neural network is parametrized by weights and biases of the network. The prediction of the neural network is compared to the data in a loss function and an optimization algorithm is used to adjust the weights and biases of a network, such that the final error is minimum. There are several optimization algorithms are available in literature, those are gradient descent, conjugate gradient and quasi-Newton etc\cite{goodfelow2016deep}. In neural network training, an epoch is a full pass over the training data. The weights, bias at each layer are determined after training. Using equation \ref{eq:14} the output $Z_l$ can be rapidly computed from the given input vector $z_0$. The computational cost of training simulations with FCNN is very less, since it involves only a few matrix multiplications.

\section{Physics based input features for machine learning}
The input features of a machine learning model should be carefully chosen for accurately defining the modelled term. Firstly this ensures that the model has the requisite information to re-create the target without over-fitting. Secondly this ensures that physics based constraints are met in the final model. For instance due to Galilean invariance we should ensure that the features in the modeling basis also obey this requirement.


In this work, the input features to the machine learning model are chosen using two different approaches, one from the modelled equation for the pressure strain correlation and another from its direct definition. The model equation for the pressure strain correlation has dependency with turbulence dissipation, Reynolds stress anisotropy, turbulence kinetic energy, strain and vorticity and the definition of the pressure strain correlation has dependency with pressure and velocity gradient. The functional mapping for the above two approaches can be written as:
\begin{equation}
\begin{split}
&\phi_{uv}=f_1(b_{uv}, \epsilon, \frac{du}{dy}, k)
\end{split}
\end{equation}
\begin{equation}
\begin{split}
&\phi_{uv}=f_2(p, \frac{du}{dy})
\end{split}
\end{equation}   
For the mapping $f_1$ we could have taken strain and vorticity separately, but for reducing the number of input parameters, in place those two, we have simply considered the velocity gradient(both strain and vorticity are related to velocity gradient). All the inputs to the neural network are normalized using the formula: $\alpha^*=\frac{\alpha-\alpha_{min}}{\alpha_{max}-\alpha_{min}}$, so that the inputs will be in the range 0 and 1. This avoids clustering of training in one direction and enhance convergence in the training. The neural network model with four input features will be termed as NN1(fig. \ref{fig:2}) and the one with two input features will be termed as NN2(fig. \ref{fig:3}) throughout the article. 

\section{Training of the neural networks}
\begin{table}
\begin{center}
 \begin{tabular}
 {||c  c c||} 
 \hline
 Case & Training Set & Testing Set \\ [0.5ex] 
 \hline\hline
 1 & $Re_\lambda=550,1000,2000$  & $Re_\lambda=5200$ \\ 
 \hline
 2 & $Re_\lambda=550,1000,5200$ &  $Re_\lambda=2000$ \\
 \hline
 3 & $Re_\lambda=550,2000,5200$ &  $Re_\lambda=1000$ \\
 
 \hline
 4 & $Re_\lambda=1000,2000,5200$ & $Re_\lambda=550$ \\ [1ex] 
\hline
\end{tabular}
\end{center}
\caption{Four training and test cases for the turbulent channel flow\cite{lee2015direct}.\label{t1}}
\end{table}

\begin{figure*}[ht]
\centering
\captionsetup{justification=centering}
\subfloat[$5200$]{\includegraphics[width=0.5\textwidth]{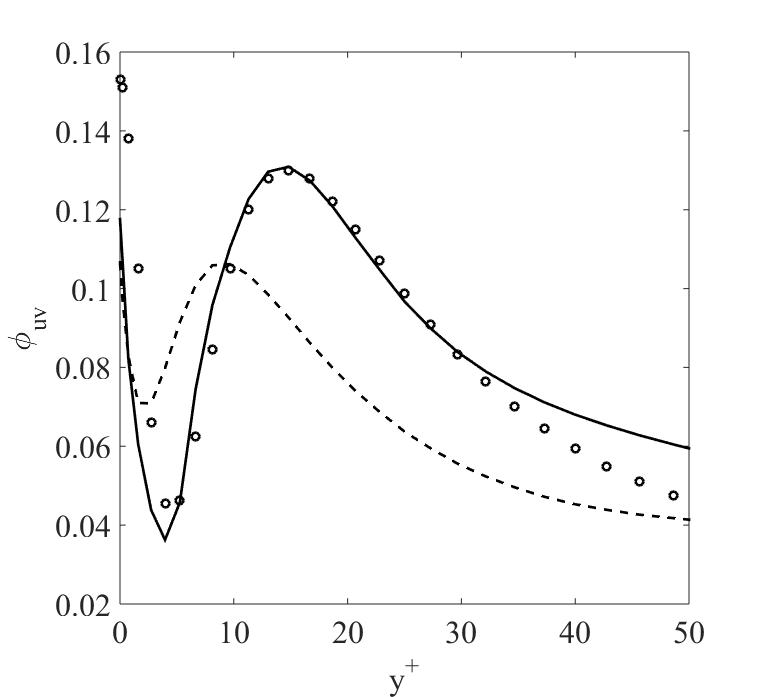}}
 \subfloat[$2000$]{\includegraphics[width=0.5\textwidth]{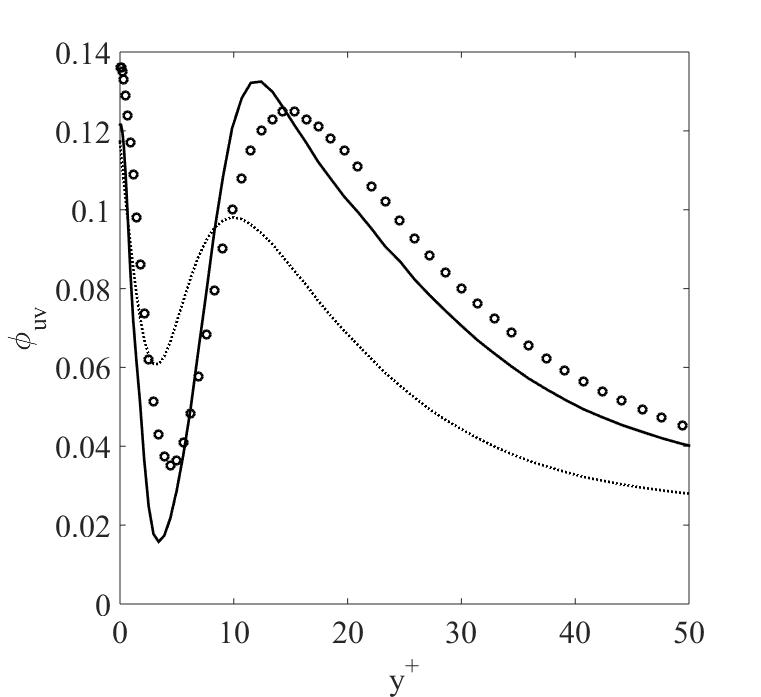}}\\
 \subfloat[$1000$]{\includegraphics[width=0.5\textwidth]{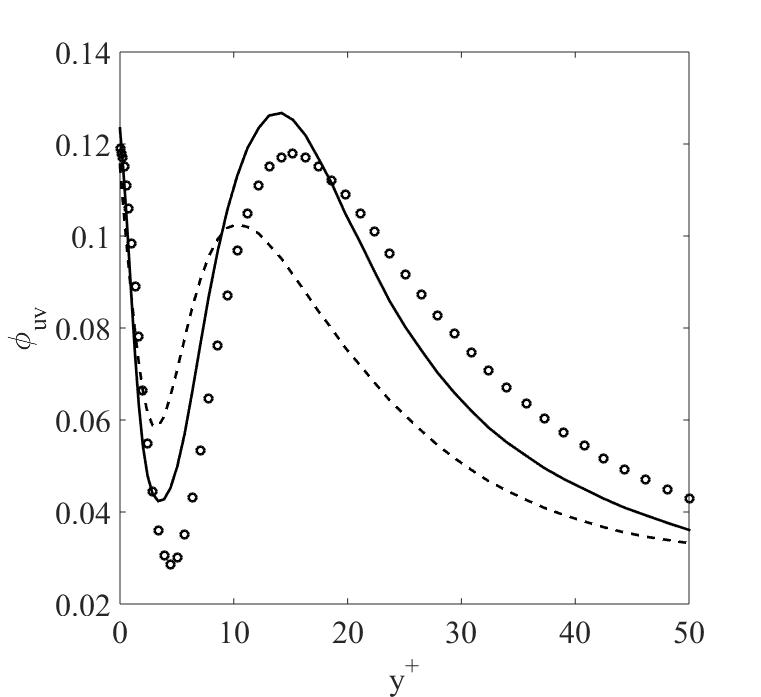}}
 \subfloat[$550$]{\includegraphics[width=0.5\textwidth]{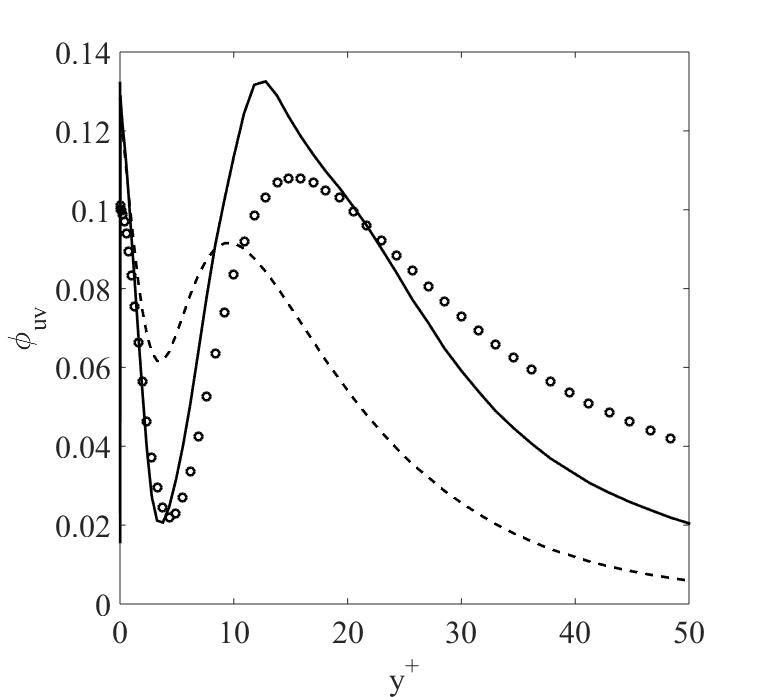}}
 \caption{Prediction of NN1 and a simple perception(SP) for the test cases (ref. table \ref{t1}). Markers(DNS), Solid line(NN1) and dashed line(SP). \label{fig:5}}
\end{figure*}
The neural network models were trained from DNS turbulent flow data from the Oden institute turbulence file server(https://turbulence.oden.utexas.edu/). The opensource library Keras\cite{chollet2015keras} with TensorFlow backend is used for training the neural network models. In the data set, mean flow profiles, turbulence statistics and the terms in the Reynolds stress transport equation are available for four different friction Reynold numbers ($Re_\lambda=550,1000,2000,5200$). The data for turbulence dissipation, turbulence kinetic energy, pressure, velocity gradient and Reynolds stress anisotropy were extracted from the data set. Based on friction Reynolds number the data were grouped into 4 different cases as shown in table 1. In each case, one friction Reynolds number data is kept for prediction/testing. for all the four cases 3 Reynolds number data were kept for training and another for testing.

For NN1, we observed best results with 1 hidden layers (with 10 neurons) and for NN2, we have taken five hidden layer with 10 neurons in each layers. The number of hidden layers and the number of neurons in each layer was chosen based the value of the correlation coefficient between the scaled outputs and the targets. For NN1 the correlation coefficient was 0.959 and for the NN2 the correlation coefficient was 0.985. We use hyperbolic tangent as non-linear activation function for the hidden layers. For optimization, we have used the Adam optimizer\cite{Goodfellow-et-al-2016}. The errors were calculated using mean squared error formula.           

\section{Testing of the trained networks}

\begin{figure*}[ht]
\centering
\captionsetup{justification=centering}
\subfloat[$5200$]{\includegraphics[width=0.5\textwidth]{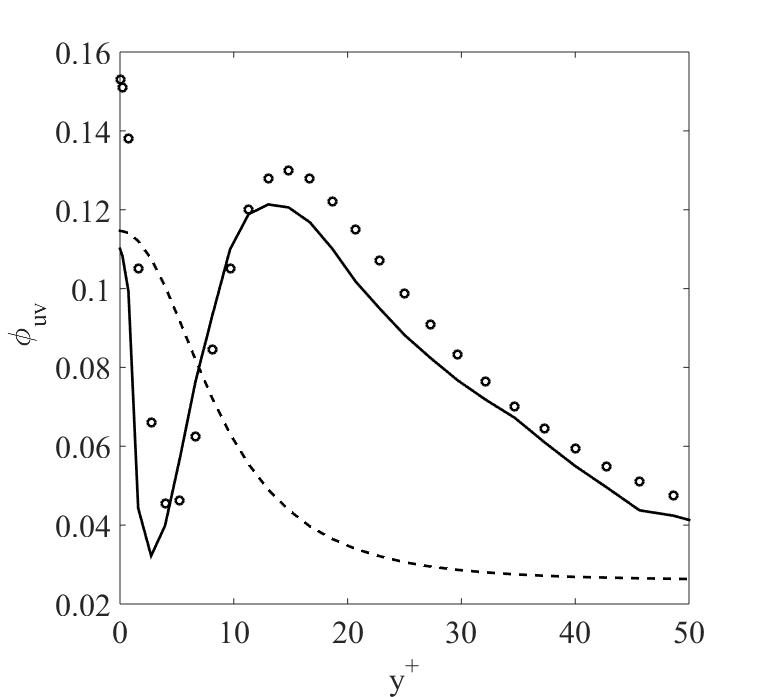}}
 \subfloat[$2000$]{\includegraphics[width=0.5\textwidth]{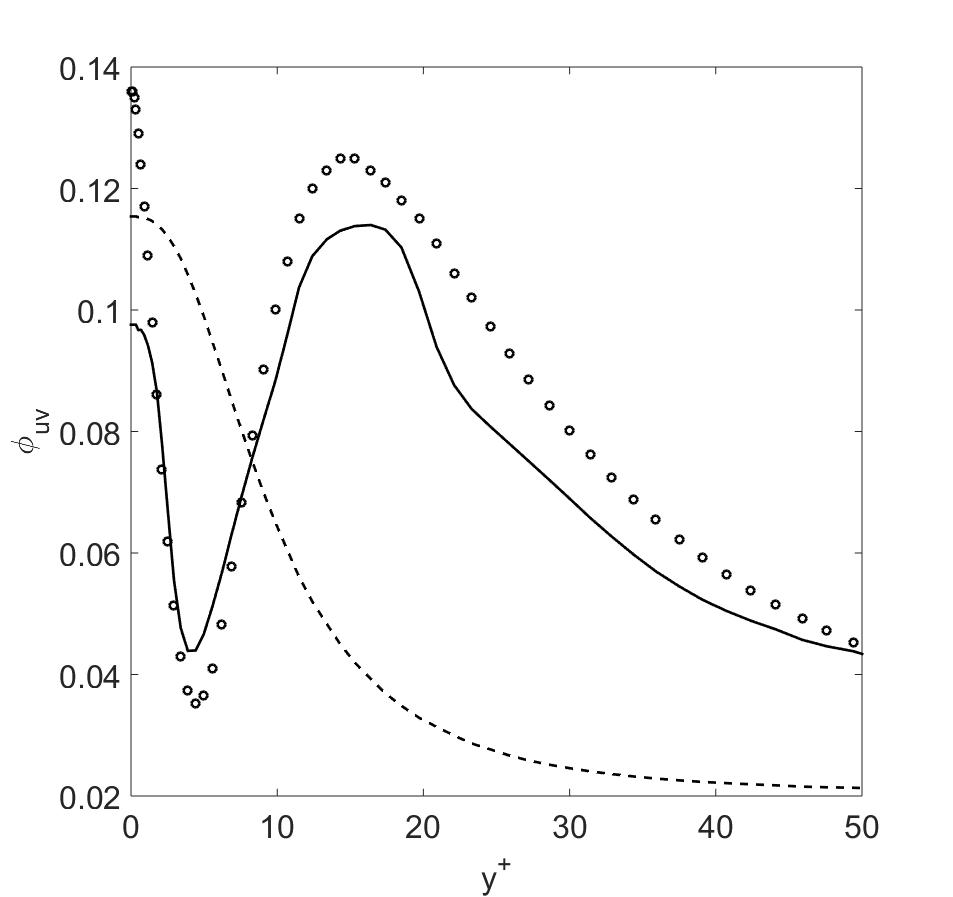}}\\
 \subfloat[$1000$]{\includegraphics[width=0.5\textwidth]{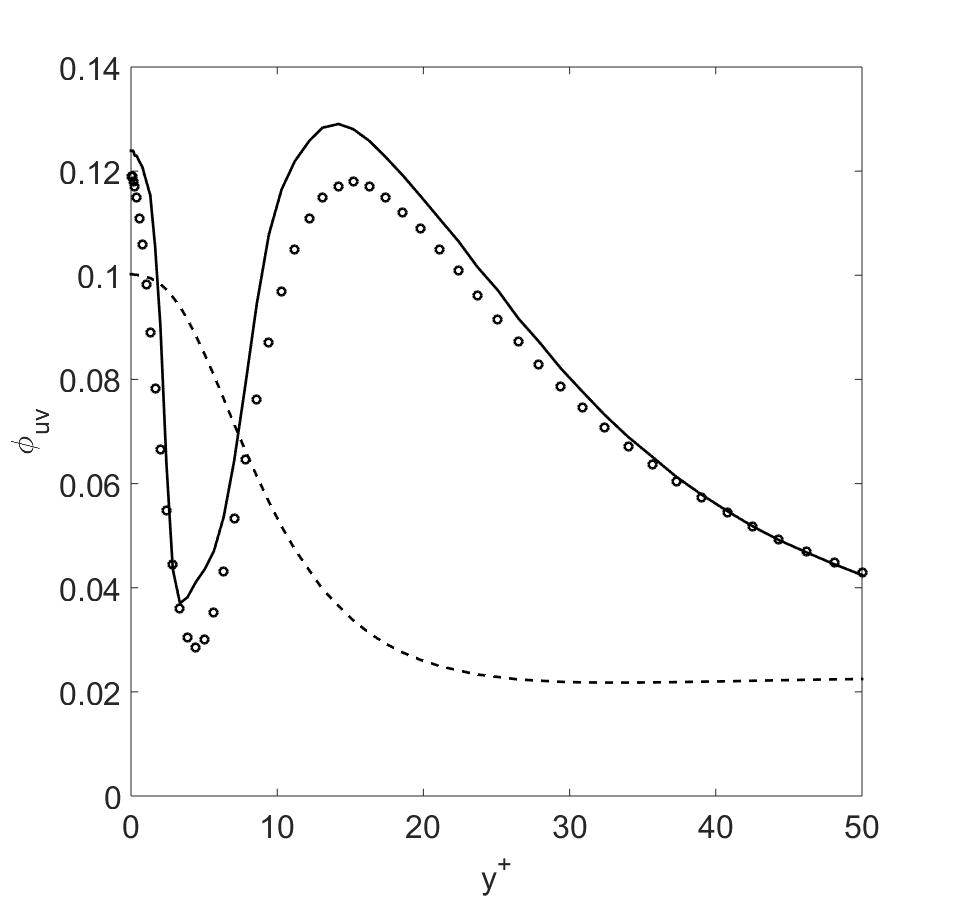}}
 \subfloat[$550$]{\includegraphics[width=0.5\textwidth]{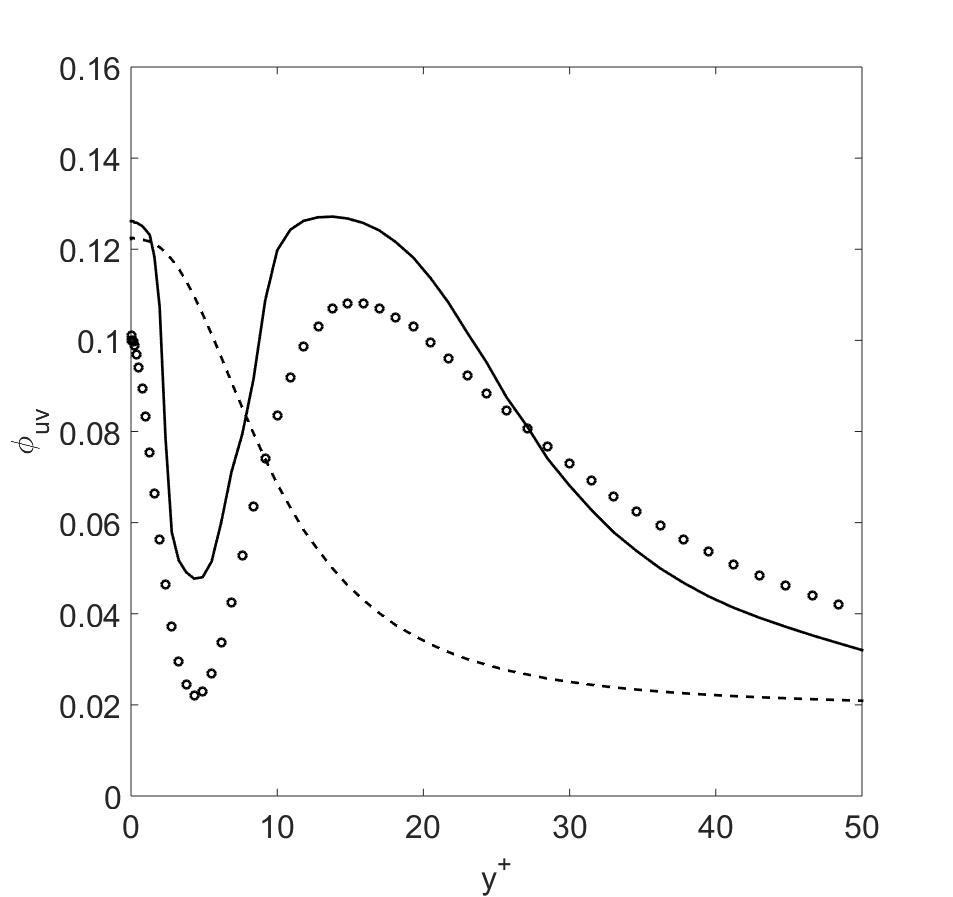}}
 \caption{Prediction of the NN2 and a SP for the test cases (ref. table \ref{t1}).Markers(DNS), Solid line(NN2) and dashed line(SP).\label{fig:6}}
\end{figure*}

After training both the networks NN1 and NN2, we have first validated the NN model predictions against the validation data. The validation data was randomly collected from the training set as shown in fig. \ref{fig:4}. The markers in fig. \ref{fig:4} represent the validation data. We have validated both the models for all $Re_\lambda$ values, but for brevity we have only presented the validation results for $Re_\lambda=5200$. As shown in the figure, the neural network models exhibit higher accuracy at corresponding lower target values of the pressure strain correlation, $\phi_{uv}$. This arises due to an asymmetry in the training data where there are more samples at lower values of the actual pressure strain correlation term. This can be corrected by weighed regression, where a corresponding higher penalty may be associated with the ranges of the measured pressure strain correlation where the sample density is lower. For the purposes of this illustrative neural network modeling we choose to retain conventional uniform penalties across samples. 

After preliminary validation, we have the tested both the models for all the testing sets from table \ref{t1}.The NN1 model predictions are shown in fig. \ref{fig:5}. The markers represent DNS data of turbulent channel flow \cite{lee2015direct}. For comparing the predictive capability of the NN1 model, simulations were also performed for a simple perception(sp). The simple perception has zero hidden layers. From fig. \ref{fig:5} it is observed that for      
$Re_\lambda=5200$ and $2000$ the NN1 model model predictions matches well with the DNS results and much better than the SP predictions. A slight discrepancy between NN1 and DNS results is observed in fig. \ref {fig:5}c and d. Here the data driven model shows a small bias in prediction and consistently overpredicts as compared to the DNS data. However the accuracy of the model is still high and it replicate the qualitative features admirably. 

The predictions of the NN2 are shown in fig. \ref{fig:6}. The input features for the NN2 were directly selected from the definition of the pressure strain correlation rather than the modelled equation. In fig. \ref{fig:6} the markers, solid lines and dashed lines represent  DNS results, NN2 predictions and SP predictions respectively. The NN2 predictions matches with the trend of DNS results but NN1 predictions are better than that of NN2. The SP completely failed to predict the pressure strain correlation. This is due to the fact that the SP model corresponds to a simple linear regression between the features and the target. This linear model does not have the flexibility to approximate the complex relationship required to model the pressure strain correlation. 

\begin{figure}
\centering
\includegraphics[width=0.5\textwidth]{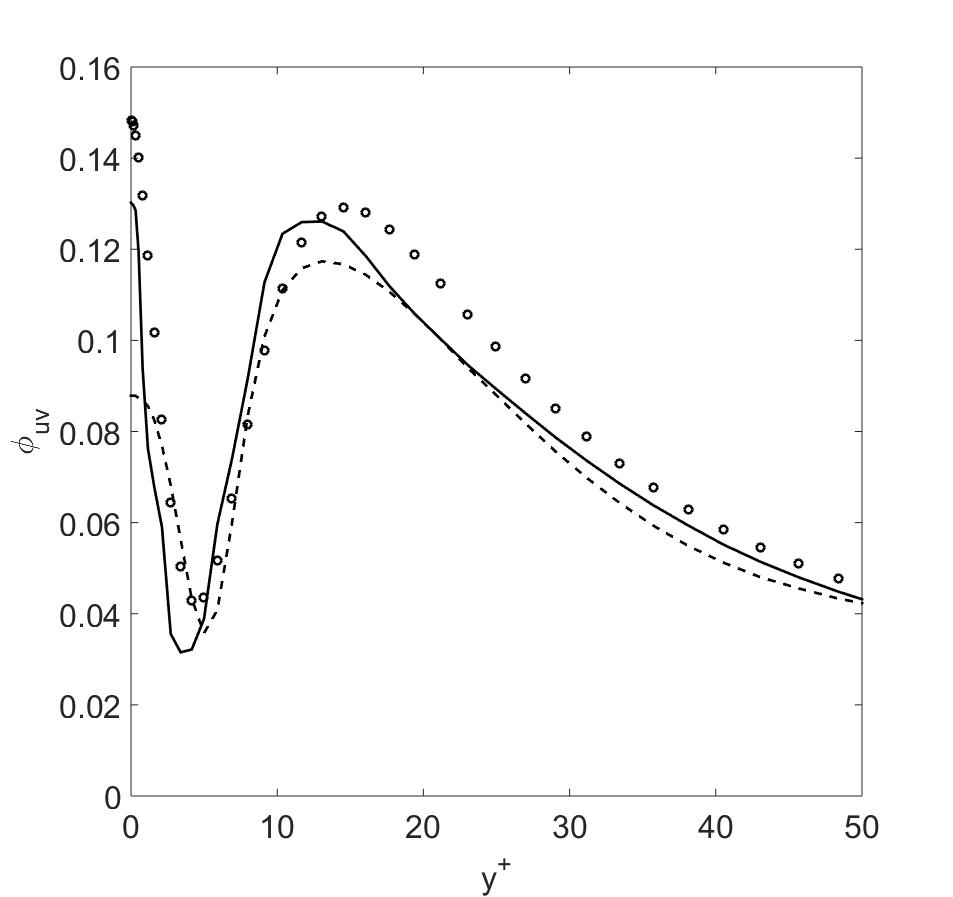}
 \caption{Prediction of NN1 and NN2 for the fully unknown test case of turbulent plane Couette flows\cite{lee_moser_2018} for $Re_{\lambda}=500$. Markers(DNS), Solid line(NN1) and dashed line(NN2).\label{fig:7}}
\end{figure}

A primary requirement for machine learning models is generalizability. The machine learning model is expected to perform well not only in the cases that were included in its training but must also perform well in similar cases that were not included in its training. To evaluate the generalizability characteristics of the trained deep learning models, we have tested the predictive capability of NN1 and NN2 for a fully unknown prediction case of turbulent plane Couette flow \cite{lee_moser_2018} at $Re_{\lambda}=500$. Both the neural networks are trained with data from case 4 of table 1.  As shown in fig. \ref{fig:7}, both the neural network predictions are matching well with the DNS results. In contrast to NN2 the predictions of NN1 are comparatively better in predicting the evolution of pressure strain. 

Thus we find that with proper selection of input features and model hyperparameters, deep learning models of the pressure strain correlation can capture qualitative trends in turbulent flow cases very well. Quantitatively the predictions of the best model are within reasonable accuracy. These deep neural network models also show good generalizability where the model performance is consistently satisfactory across similar turbulent flow cases that were not utilized in the training of the model. 

\section{Concluding remarks}

Machine learning based approaches are being rapidly utilized in different domains of physics including turbulence modeling. In turbulence modeling the application of these data driven approaches is limited to eddy viscosity based turbulence closures. The core thesis of this investigation is that the limited expressivity of such two-equation models delimits the information that they can utilize from the data. We outline with detailed arguments the manner in which the eddy-viscosity hypothesis, the instantaneous linear relationship between rate of stain and the Reynolds stresses, the inability to resolve high degrees of turbulence anisotropy encumber the potential of machine learning models. We propose that the Reynolds Stress Modeling approach may be a more appropriate level of closure for the application of data driven modeling.  

As an illustration, deep neural network models were developed for pressure strain correlation of turbulence considering DNS data of turbulence channel flow at different friction Reynolds numbers. Two different input feature sets were considered for the networks, one based upon the modelled equation and another based upon the actual definition of the pressure strain correlation. In contrast to NN2, the NN1 model predictions matches well with the DNS results. However there is little discrepancy between the DNS results and NN1 model predictions at lower Reynolds numbers. Such discrepancy could be improved by incorporating information about wall, Reynolds number or non-local effects into input feature space of the model or by modifying the structure of the neural network. For example non-local information can be integrated into the modeling framework naturally by the inclusion of convolutional layers in the neural network architecture. The proposed models can be incorporated into CFD codes as source terms of the Reynolds stress transport equations in conjunction with the models for the normal and shear components of the pressure strain correlation. 

\bibliographystyle{SageV}
\bibliography{asme2e}

\end{document}